\documentclass[twocolumn,nopacs,preprintnumbers]{revtex4}
\usepackage{graphicx}
\usepackage[francais]{babel}
\usepackage[latin1]{inputenc}
\usepackage{bm}
\usepackage{color}
\usepackage[normalem]{ulem}

\usepackage{physics}
\usepackage{subfigure}

\graphicspath{%
    {converted_graphics/}
    {/}
}
\begin{document}

\title{Quantum control beyond the adiabatic regime in 2D curved matter-wave guides}


\author{Fran\c{c}ois Impens$^1$, Romain Duboscq$^2$ and David Gu\'ery-Odelin$^3$}
\affiliation{$^1$ Instituto de F\'{i}sica, Universidade Federal do Rio de Janeiro,  Rio de Janeiro, RJ 21941-972, Brazil
\\
$^2$ Universit\'e de Toulouse ; CNRS, INSA IMT, F-31062 Toulouse Cedex 9, France 
\\
$^3$ Laboratoire Collisions, Agr\'egats, R\'eactivit\'e, IRSAMC, Universit\'e de Toulouse, CNRS, UPS, France 
}


\date{\today}

\begin{abstract}
The propagation of matter waves in curved geometry is relevant for electrons in nano-wires, solid-state physics structures and atomtronics. Curvature effects are usually addressed within the adiabatic limit and treated via an effective potential acting on the manifold to which the particles are strongly confined. However, the strength of the confinements that can be achieved experimentally are in practice limited, and the adiabatic approximation framework often appears too restrictive for the realistic design of relevant propagation structures. Here, we explore the design of 2D sharply bent wave-guides for the propagation of matter waves beyond the adiabatic regime. The bend design, which enables the connection of guide components rotated by an arbitrary angle and of arbitrary curvatures, rests on an exact inverse-engineering technique. The resolution of the full 2D Schr\"odinger equation in curved geometry shows that  our method yields reflectionless guides with a transverse stability improved by several orders of magnitude when compared to circular guides of similar size. 
\end{abstract}

\maketitle

\textit{Introduction - }  The development of quantum technologies builds up on the increasing level of control of both internal and external degrees of freedom of atoms. This has motivated the design of quantum control protocols as key ingredients for such technologies. Quantum control protocols are mostly developed for quantum observables associated to flat coordinate systems. However, the geometry can drastically affect the quantum properties~\cite{Costa1,Costa2}: for instance, any kind of bending produces a bound state ~\cite{Goldstone92}. This suggests that geometry provides a powerful tool to control quantum systems. For instance, new form of wave localization have been proposed by combining curved potentials and topological protection~\cite{ArXivAAHgrating2020}. 

Since the pioneering works of Costa \cite{Costa1,Costa2} and, Goldstone and Jaffe \cite{Goldstone92} to establish the framework of quantum mechanics in curved geometry, the literature has focused on an effective 1D treatment relying on an adiabatic approximation \cite{nano1,nano2,Leboeuf01,nano3,DelCampo14,nano4,nano5}. In this approach, the effect of curvature is encapsulated in an effective attractive potential~\cite{Goldstone92}.  However, as explained below, the validity of quantum control protocols based on such a treatment is restricted to regimes of weak and slowly-varying curvatures. Furthermore, many experimental situations such as  electrons in quantum nano-wires~\cite{nt1,nt2,nt3,ElecCurv1,ElecCurv2,Cheng2020} or atomic wave guides \cite{AtomtronicsNJP17} often require to go beyond this 1D approximation. In this letter, we consider the real 2D problem and set up a non-adiabatic quantum control strategy valid in strongly curved geometry. 
 
Such a study is relevant for the growing field of nanomaterials having complex geometries. This includes corrugated carbon nanotubes \cite{nt1}, rolled-up nanotubes \cite{nt2}, and M\"obius nanostructures \cite{nt3} to name a few. In the field of atomtronics, many techniques have been investigated to design guides including static and/or time-modulated magnetic trapping, off-resonant light and rf-induced trapping  \cite{R1,R2,R3,R4,R5,wolf}. The control of the external degrees of freedom has also undergone an extraordinary progress with the realization of atom lasers having a high quality factor ~\cite{Ketterle02,Guerin06,Riou06,Impens08,Impens09b} or/and placed in the transverse ground state~\cite{Close07,AtomLaser1,Gattobigio09}. Guides with various shapes including rings have already been demonstrated \cite{R1,R2,R3,R4,R5,R6,R7,R8,R9,wolf,Reichel01,cornell}. The applications and the miniaturization of matter wave circuits require a perfect control of matter wave propagation in a bent guide. 
 A sharp bending favors compactness at the expense of a coupling between longitudinal and transverse degrees of freedom. This coupling may have deleterious consequences for the control of the matter wave as it may become chaotic \cite{dgo1,dgo2}.

In the following, we first work out an inverse engineering strategy, inspired by shortcuts to adiabaticity (sta) protocols~\cite{Shortcut1,Shortcut2,RMPSTA19}, to shape a class of classical trajectories robust to a variation in the initial conditions and free of residual transverse excitations after a bend. We then validate those solutions by a numerical resolution of the full 2D Schr\"odinger equation in curved space.  Our findings reveal how a proper guide design can dramatically reduce the transverse excitations after the bend, when compared to  circular guides of identical radius and stiffness.

\textit{Schr\"odinger equation in curved geometry --}
The coordinate of a material point using curvilinear coordinates $(s,y)$ associated to a path $\mathbf{r}_c(s)$ are given by $\mathbf{r}(s,y)=\mathbf{r}_c(s) + y \mathbf{n}(s)$ where $\hat{\mathbf{n}}(s)$ is the local normal to the path. The variation with the curvilinear coordinate of the local normal vectors reads: $d \hat{\mathbf{t}} / ds = \kappa \hat{\mathbf{n}}$ and $d \mathbf{n} /ds = - \kappa \hat{\mathbf{t}} $ where $\hat{\mathbf{t}}(s)= d \mathbf{r}_c /ds$ is the local tangent and $\kappa(s)=1/R(s)$ the local curvature.  

The quantum mechanical wave-function in curvilinear coordinates $(s,y)$ can be written as $\psi(s,y) =h^{-1/2}(s,y) \tilde{\phi}(s,y).$  The complex-valued function $\tilde{\phi}(s,y)$ is given by a direct coordinate change in the cartesian coordinates  wave-function $\phi(X,Y)$: $\tilde{\phi}(s,y) = \phi(X(s,y),Y(s,y))$ where $(X(s,y),Y(s,y))$ are the cartesian coordinates associated to the classical point $\mathbf{r}(s,y) $ defined above. The factor $h^{-1/2}(s,y)$  arises from the Jacobian associated to the passage from cartesian to curvilinear coordinates. The bending of the guide causes indeed a local variation of the metric, captured by the definition of a function $h(s,y)=1-\kappa(s)y$ depending on the local path curvature $\kappa(s)$. Similarly to general relativity, this inhomogeneous metric encodes the inertial forces. It is also responsible for the coupling between longitudinal to transverse degrees of freedom. 
 
In its most general form, the curvilinear wave-function $\psi(s,y)$ satisfies the following time-dependent Schr\"odinger equation in the presence of a transverse confining potential $V_{\perp}(y)$~\cite{Goldstone92}:
\begin{eqnarray}
& & \left[ - \frac {\hbar^2} {2 m} \left( \frac {1}  {h(s,y)} \frac {\partial} {\partial y}  h(s,y) \frac {\partial} {\partial y}+ \frac {1}  {h(s,y)} \frac {\partial} {\partial s} \frac {1}  {h(s,y)} \frac {\partial} {\partial s} \right) \right. \nonumber \\
 & \: & \qquad \qquad \qquad \qquad +  \left. V_{\perp}(y) \frac {} {}  \right] \psi(s,y) = i \hbar \frac {\partial \psi(s,y)} {\partial t} \label{eq:curved Shrodinger}
\end{eqnarray}

To recover the widely used adiabatic approximation, three criteria shall be fulfilled:
\begin{equation}
\label{eq:adiabaticity}
 ({\rm a}) \: \: \sigma |\kappa| \ll 1, \quad  ({\rm b}) \: \: \sigma \left| \frac {d \kappa}  {ds} \right| \ll \kappa, \quad  ({\rm c})  \: \:  \sigma \left| \frac {d^2 \kappa}  {ds^2} \right| \ll \kappa^2 ,
\end{equation}
where $\sigma$ is the transverse size of the wave-packet. In this limit, the Schr\"odinger equation becomes separable and yields independent longitudinal and transverse motions. 
Within the adiabatic approximation, the effect of curvature is encapsulated in an effective attractive 1D potential $V_{\rm eff}(s)=-\hbar^2 \kappa(s)^2/(8m)$, valid for sufficiently strong transverse confinement.

\begin{figure}[htbp]
    \centering
    \includegraphics[width=4 cm]{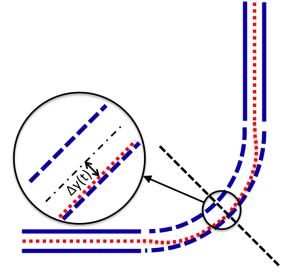}
    \caption{Problem statement: How to connect two straight guides without inducing extra transverse oscillations in the output guide (trajectory represented by a dotted red line) ?}
    \label{fig:1}
\end{figure}

Our aim is to design the shape of a guide that connects two guides. To fix ideas, we consider a relative angle between the guides equal to $\alpha=90^{\circ}$, the same transverse confinement for each guide, $V_{\perp}(y) = \frac 1 2 m \omega^2 y^2$, and an initial and final guide straight (see Fig.~\ref{fig:1}). The generalization to initial and final curved wave guides and for another relative angle is straightforward. 
The simplest candidate is a quarter-of-circle of constant radius $R$ connecting the two straight guides. However, the abrupt change of curvature at the entrance and exit of such a bend generates a sudden centrifugal force that induces transverse excitations. To circumvent this limitation, we show in the following how to tailor the curvature profile $\kappa(s)$ as a function of the curvilinear coordinate $s$. At first sight, the direct solution of this problem through Eq.~(\ref{eq:curved Shrodinger}) seems very challenging. We overcome this difficulty by using a class of properly tailored classical solutions.

\textit{Exact inverse engineering --}  The Newton law expressed in curvilinear coordinates yields the following set of two coupled nonlinear equations:
\begin{eqnarray}
\ddot{s} (1-\kappa y) - \dot{s} (\dot{\kappa} y + 2 \kappa \dot{y}) & = &0 \label{eq:dynamics_first} \\
\ddot{y} + \omega^2 y   +  \dot{s}^2 \kappa (1 -\kappa y) & = & 0  \label{eq:dynamics_second}
\end{eqnarray}
with $\kappa \equiv \kappa(s(t))$.
Combining Eqs.~(\ref{eq:dynamics_first}) and (\ref{eq:dynamics_second}), we recover the conservation of energy:
\begin{equation}
\label{eq:constantofmotion}
\dot{y}^2 + \omega^2 y^2   +  v_{ \kappa}^2  = {\rm Cte} = \frac {2E} {m},
\end{equation}
where   $v_{\rm \kappa}=\dot{s}(1-\kappa y)$ ($\dot{v}_{ \kappa}= \dot{s} \kappa \dot{y}$ from Eq.~(\ref{eq:dynamics_first})).
The quantity $K_{ \kappa}(\dot{s},s,y)= \frac 1 2 m v_{\kappa}^2 $ is nothing but the longitudinal kinetic energy in a straight guide, and encapsulates also the potential induced by inertial forces in a curved guide.
An exact inverse engineering of the transverse motion can be worked out by imposing the desired smooth trajectory for the transverse coordinates, $y(t)$. From Eq.~(\ref{eq:constantofmotion}), we then get $v_{\kappa}(t)$ from $y(t)$. This latter quantity gives access to both  the longitudinal velocity from 
$\dot{s} = v_{\kappa}  + \dot{s} \kappa y= v_{\kappa}  + \dot{v}_{\kappa}  y / \dot{y}$ and the time-dependent curvature at the particle position $\kappa(s(t))= \dot{v}_{\kappa}/ \dot{s}(t) \dot{y}(t)= \dot{v}_{\kappa}/ (d (v_{\kappa} y)/dt)$. The curvature profile $\kappa(s)$ is eventually reconstructed by integration of the longitudinal velocity.
 
For the connection between two orthogonal straight guides (see Fig.~\ref{fig:1}), we exploit the symmetry (dashed line at 45$^\circ$) and proceed in two symmetrical steps. The connection is performed in a total time $2T$. To design the guide on the first segment of duration $T$, we impose the following boundary conditions: $y_{\rm sta}(0)=0$, $y_{\rm sta}(T)= \Delta y$, $\dot{y}_{\rm sta}(0)=\dot{y}_{\rm sta}(T)=0$ and $\ddot{y}_{\rm sta}(0)=\ddot{y}_{\rm sta}(T)=0$. The conditions at $t=0$ translate the absence of transverse excitation at the entrance of the bend. The conditions at $t=T$ are required for two reasons: it ensures the continuity of the position and velocity in the middle of the guide and it enforces the stability of the trajectory against the final time $2T$, and as such it improves the resilience against a small dispersion in the longitudinal velocity \cite{RMPSTA19}. This latter requirement is relevant for the propagation of a matter wave since the wave packet has finite size and thus a finite velocity dispersion. To accommodate for the boundary conditions, we choose a trajectory in the form of a polynomial $y_{\rm sta}(t)=P(t/T)$ with $P(x)=\Delta y(10 x^3-15 x^4+6 x^5)$. Many other choices could be made for this interpolating function depending on the relevant constraints as discussed in \cite{transportdgo,dgosugny}. 

The free parameter $\Delta y$ is fixed by the choice of the maximum curvature, $\kappa_m$, reached at $t=T$: $\Delta y = -\frac 1 4 \kappa_m^{-1} \left( \sqrt{1+8 \dot s(T)^2 \kappa_m^2/ \omega^2} -1 \right)$. The time $T$ is determined self-consistently by the choice of the rotation angle $\alpha' = \langle \hat{\mathbf{t}}(s(0)), \hat{\mathbf{t}}(s(T))  \rangle= \int_0^{T} dt \:  \kappa(t) \dot{s}(t)$ ($\alpha' = \pi/4$ for our example):
\begin{equation}
\alpha'=  - \int_0^T dt \frac{\ddot{y}_{\rm sta}(t)+\omega^2 y_{\rm sta}(t) }{ \sqrt{2 E /m -\dot{y}^2_{\rm sta}(t)-\omega^2 y_{\rm sta}^2(t) }}.
\end{equation}
By symmetry, the solution for $T \leq t \leq 2T$ reads: $y_{\rm sta}(t) = P(1-t/T)$.

\begin{figure}[htbp]
    \centering
    \includegraphics[width=8.5 cm]{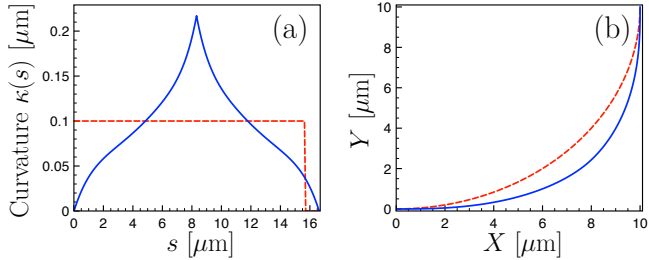}
    \caption{(a) Inverse-engineered curvature profile $\kappa(s)$ (solid blue line) and the corresponding effective circular bend of radius $R_{\rm eq} = 10 \: {\rm \mu m}$ (red dashed line) as a function of the longitudinal coordinate $s$. (b)  View from above of the inverse-engineered path (blue) shape and of the equivalent circular path (red). Parameters: initial longitudinal velocity of $\dot{s}_0= 20 \: {\rm mm/s} $, trapping frequency $\omega= 2 \pi \times 1705~{\rm Hz}$, and a maximum curvature $\kappa_m= 0.22~{\rm \mu m^{-1}}$. The total length is $s_{\rm f} \simeq 16.6 \: {\rm \mu m} $ corresponding to a total time $2 T \simeq 0.88 \: {\rm ms} $. }    \label{fig:2}
\end{figure}

 To quantify the compactness of the tailored bend, we define the effective radius of curvature $R_{\rm eq}$ as the radius of the largest quarter-of-circle enclosed within the engineered bend. 
Figure~\ref{fig:2}a gives an example of engineered curvature profile with $R_{\rm eq}= 10 \: {\rm \mu m}$ for typical experimental parameters  $\dot s_0= 20 \: {\rm mm/s} $ and $\omega/2\pi= 1705~{\rm Hz}$ \cite{Ryu15}. Figure~\ref{fig:2}b shows the engineered bend together with the effective radius of curvature $R_{\rm eq}$.

\textit{Robust cancellation of the transverse excitations -} We now compare the transverse stability achieved by the engineered vs circular bends. For a bend defined by a quarter-of-circle, the curvature is a step function that drives transverse oscillations that generally persist beyond the bend. The amplitude of these oscillations depends on the particle parameters at the entrance of the guide and on the phase of the transverse oscillation at the exit of the bend. As a result, there exist a set of discrete radius values for which there is an exact cancellation of the oscillations of the outgoing particle. 

\begin{figure}[htbp]
    \centering
    \includegraphics[width=7.5 cm]{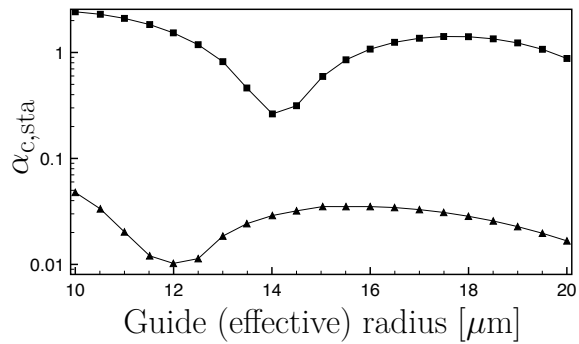}
    \caption{Averaged oscillation amplitudes $\overline{\alpha}_{\rm sta}$ for the engineered guide  (triangle) and  $\overline{\alpha}_{\rm c}$  for the circular guide (square) as a function of the guide radius. Same parameters as for Fig.~\ref{fig:2}. }    \label{fig:3}
\end{figure}

Nevertheless, even in these most favorable circumstances, it turns out that the 2D inverse-engineered guide outperforms the circular path when one considers the average transverse stabilisation achieved over a finite velocity interval.  
As already explained above, a desirable feature of a matter wave guide is indeed to suppress transverse excitations over a finite range of longitudinal velocities and not only for isolated values. To quantify the robustness of the guide, we consider an uniform distribution of initial velocities over the interval $[(1-\epsilon) \dot{s}_0,  \: (1+\epsilon)  \dot{s}_0] $, where  $\dot{s}_0$ is the incident longitudinal velocity used in the inverse-engineering procedure and with $\epsilon=5\%$. Each initial velocity $v$ yields oscillations at the exit of the bend of finite amplitude $a_{\rm c}(v)$ and $a_{\rm sta}(v)$  for the circular and the inverse-engineered guide respectively. It is instructive to compare these amplitudes to a relevant scale length for the atomic wave-packet propagation. For this purpose, we introduce the dimensionless quantities $\overline{\alpha}_{\rm c,sta} = ( 2 \epsilon \dot{s}_0 \sigma )^{-1} \int_{\dot{s}_0-\epsilon}^{\dot{s}_0+\epsilon}  \: a_{\rm c,sta}(v)dv,$ corresponding to the ratio between the averaged oscillation amplitudes and the width $\sigma=\sqrt{\hbar / m \omega}$ of the transverse ground state wave-function. For an oscillation amplitude $a$, the outgoing particle has a transverse mechanical energy $E$ which compares to the energy quantum as $E/\hbar \omega= \frac 1 2 (a/\sigma)^2$. Figure \ref{fig:3} shows the oscillation amplitudes $\overline{\alpha}_{\rm c,sta} $ as a function of the effective radius. The engineered guide yields transverse excitations that are at least one order of magnitude (and for some radii more than two orders of magnitude) smaller than that of the circular guide. The minimum of $\overline{\alpha}_{\rm c}$ results from a phase matching condition as explained above. The same effect occurs for the inverse engineered guide for a similar guide length.

\textit{Comparison between 2D and 1D curvature designs -}To facilitate the comparison between our 2D treatment and the 1D effective approach \cite{nano1,nano2,Leboeuf01,nano3,DelCampo14,nano4,nano5} we define in the following manner the 1D-adiabatic design. 

The adiabatic limit yields an independent longitudinal motion driven by an attractive bending potential $V_{\rm eff}(s)$. The longitudinal velocity is thus expected to increase with the local curvature as $\dot{s}_{\rm cl} = \sqrt{\dot{s}_0^2- 2V_{\rm eff}(s_{\rm cl})/m}$ with $\dot{s}_0$ the longitudinal velocity of the atomic wave-packet before the bend. From the classical equation (4) we infer the expression of the curvature $\kappa_{\rm 1 D}(s)$ as a function of the transverse trajectory  $y_{\rm sta}(t)$ defined previously with $T=T_{\rm 1D}$ whose value is determined by the desired rotation angle $\pi/4=\int_0^{T_{\rm 1D}} dt \:  \dot{s}_{\rm cl} (t) \kappa_{\rm 1 D}(s_{\rm cl}(t))$. We use this determination of the curvature in 2D simulations performed either with classical equations or the Schr\"odinger equation (1).

The 2D-curvature design is obtained from the procedure explained in the previous section. It relies on a similar transverse trajectory $y_{\rm sta}(t)$ but defined with the time $T_{\rm 2D}$ to get the desired angle (see (6)).

\begin{figure}[htbp]
    \centering
    \includegraphics[width=9 cm]{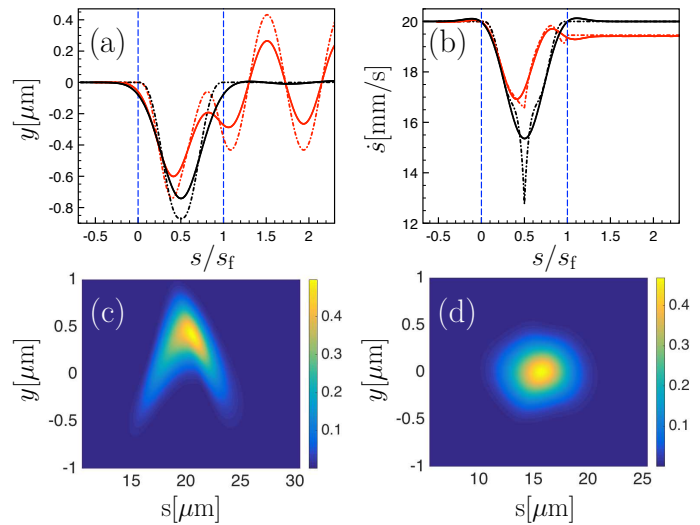}
    \caption{(a) and (b): Transverse position, $\langle y(s / s_{\rm f}) \rangle$, and longitudinal velocity, $\langle \dot{s}\rangle(s / s_{\rm f})$, of the wave packet as a function of the normalized longitudinal coordinate, $s/s_{\rm f}$. The curves are obtained from 2D classical simulations using either the 1D-adiabatic design (red dot-dashed line) or the 2D-optimized protocol (black dot-dashed line), or from the 2D Schr\"odinger equation~(\ref{eq:curved Shrodinger}) using either the 1D-adiabatic design (red solid line) or the 2D-optimized protocol (black solid line). 
Color plot of the modulus square of the 2D wave-function $|\phi(s,y)|^2$ (in ${\rm \mu m}^{-2}$) after propagation at the position $s=1.5 s_f$ for the 1D-adiabatic design (c) and 2D-nonadiabatic design (d). Parameters: initial Gaussian wave-function of widths $\sigma_y=\sigma_s/10=(\hbar/m\omega)^{1/2}$ centered on axis ($y=0$) at $s =-0.5s_{\rm f}$ and with a velocity $\dot{s}_0=20$ mm/s.}
  \label{fig:4}
\end{figure}

We work out here a concrete example with the parameters: $\dot{s}_0=20$ mm/s and $\omega= 2 \pi \times 1705~{\rm Hz}$. We find  $T_{\rm 1D} = 0.334 \: {\rm ms}$ for $\kappa_{\rm 1 D}(s)$ and $T_{\rm 2D} = 0.295 \: {\rm ms}$ for $\kappa_{\rm 2 D}(s)$, corresponding to the respective bend lengths $s_{\rm f \: 1D}=13.36 \: {\rm \mu m} $ and $s_{\rm f \: 2D}=10.37 \: {\rm \mu m}$. The initial wave-function is a cigar-shaped Gaussian wave-function of widths $\sigma_y=\sigma_s/10=(\hbar/m\omega)^{1/2}$ initially centered on the axis $y=0$, at the normalized longitudinal position $s/s_{\rm f}=-0.5$ and with an average velocity $\dot{s}_0$ (see Fig.~\ref{fig:4}). With this initial position, the wave packet is  at $t=0$ entirely outside the bent. 

 The effectiveness of the two methods is analyzed through both the classical 2D Newton equations and the numerical resolution of the full 2D Schr\"odinger equation~\eqref{eq:curved Shrodinger} for both curvature profiles $\kappa_{\rm 1 D}(s)$ and $\kappa_{\rm 2 D}(s)$. The quantum simulation is based on a Crank-Nicolson scheme. The results are summarized on Fig.~\ref{fig:4} where the transverse position and longitudinal velocities of the packet are plotted as a function of the normalized longitudinal coordinate $s/s_{\rm f}$ for both curvature designs. 

It is worth noticing that for our choice of parameters the curvature $\kappa(s)$ varies considerably on the width of the wave-packet as $\sigma_s / s_{\rm f} \sim 0.2$ for both guides. This contributes to the clear difference between the classical (dot-dashed lines) and quantum trajectories (solid lines).  With the 1D adiabatic-design bend, we observe strong transverse oscillations that persist in the output straight guide  ($s / s_{\rm f}\geq 1)$ (see Fig.~\ref{fig:4}a). This design therefore fails to provide a reliable connection between the two straight guides for both classical and quantum simulations. A totally different behavior is obtained with the 2D-curvature design: the transverse oscillations in the output channel are almost completely suppressed. As a figure of merit, we evaluate for each guide the excess of transverse energy of the wave-packet once in the output straight guide with respect to the ground state energy scale $\Delta E_{\rm t.}= \langle \frac {1} {2m} p_y^2+  \frac {1} {2} m \omega^2 y^2  \rangle -  \frac 1 2 \hbar \omega = \overline{n} \hbar \omega$. We obtain the respective average number of transverse excitations quanta $\overline{n}_{\rm 1D}=1.4$ and $\overline{n}_{\rm 2D}=5.1 \times 10^{-3}$ for the 1D-adiabatic and 2D designs respectively. With our 2D protocol, we reach a fidelity up to 99.6 \%. These results validates our strategy for well-controlled matter-wave propagation in curved guides. 

Figure~\ref{fig:4}b shows the evolution of the longitudinal velocity along the bend for both guides. For the 1D-guide, the final longitudinal velocity obtained from quantum simulations is noticeably below its initial value, which witnesses the transfer of energy between the longitudinal and transverse degrees of freedom. A common feature of both bent guides is that the longitudinal velocity significantly decreases when the wave-packet is expelled from the bend center. This is a fingerprint of angular momentum conservation, even though with the considered position-dependent curvature profile angular momentum is not rigorously conserved.  This correlation is a signature that the propagation occurs beyond the adiabatic regime, for which the longitudinal and transverse motion are expected to be independent~\cite{Goldstone92,nano1,nano2,Leboeuf01,nano3,DelCampo14,nano4,nano5}.  It is remarkable that the longitudinal velocity obtained from the 2D Schr\"odinger equation is indeed minimal in the region of strong curvature, where it should be maximal according to the 1D adiabatic approach. This suggests that the convergence of the longitudinal velocity profile towards the adiabatic limit is slow.  Indeed the considered 2D curvature profile $\kappa_{\rm 2D}(s)$ presents rapid variations which violate the adiabatic criteria~(\ref{eq:adiabaticity}b) and ~(\ref{eq:adiabaticity}c), as $\sigma |d \kappa_{\rm 2D} / ds| \simeq 0.6 \kappa$ and $\sigma |d^2 \kappa_{\rm 2D} / ds^2| \simeq 5 \kappa^2$ in the vicinity of the strongly curved region of the bend. In this regime, our procedure clearly outperforms previous methods based on an effective 1D potential. Note that our method only breaks down when the condition \ref{eq:adiabaticity}(a) is violated. Indeed, in this extreme limit, the classical/quantum correspondence fails down since the inertial force strongly varies over the wave-packet size.   
  
 To conclude, we have presented a systematic procedure to design reflectionless strongly curved 2D matter wave guides with unprecedented fidelity.  
 Our study has revealed the weakness of previous studies based on 1D effective potential, and the fact that the convergence of a real system towards such an adiabatic limit turns out to be very slow. Our approach provides a new strategy for quantum control in curved geometry beyond the adiabatic regime, and is a step forward towards the realization of highly stable complex matter wave circuits.
 Future prospects for this work include the propagation in curved geometry in the presence of spin-orbit interactions \cite{soi}.
  
\acknowledgements
This work is part of INCT-IQ from CNPq. It was supported by the Brazilian agencies CNPq, CAPES, and FAPERJ and the Agence Nationale de la Recherche research funding Grant No.~ANR-18-CE30-0013.

\end{document}